\documentstyle[preprint,aps,floats,psfig]{revtex}
\tighten
\begin{document}
\title{Further evidences for a generic Universe}
\author{C\'esar A.\ Terrero-Escalante, Alberto A. Garc\'{\i}a}
\address{Departamento~de~F\'{\i}sica,
~Centro de Investigaci\'on y de Estudios Avanzados del IPN,
~Apdo.~Postal~14-740,~07000,~M\'exico~D.F.,~M\'exico.}
\maketitle

\begin{abstract}
Recently it was argued that an inflationary potential yielding power spectra 
characterized by a scale-invariant tensorial spectral index and a 
weakly scale-dependent scalar spectral 
index might account for a generic large-scale structure formation in the 
multiverses scenario of eternal inflation. Here it is shown that this
statement remains true if the tensorial index is allowed to slowly vary in
a wide range of angular scales. In the cases analyzed in this paper, the scalar
index also slightly depends on the scales. Therefore, the production of closely
resembling each other universes seems to be a common feature of inflationary
scenarios with weakly scale-dependent spectral indices. With power-law 
inflation as the extreme case, this is the class of inflationary
scenarios which may give the best fit to near future CMB observations. 
\end{abstract}

\pacs{PACS numbers: 98.80.Cq, 98.80.Es, 98.70.Vc}

\section{Introduction}
\label{sec:Intro}

In this paper we would like to address the question of how probable is
the existence of a Universe like this that we can observe. Obviously, the 
answer to this question must integrate the role of a large number of 
factors but
here we shall focus in the generic description of our Universe from the point
of view of matter structures we can observe in it. Also, in our
analysis we shall argue neither in favor nor against the anthropic principle.
Thus, to answer our question it is necessary, first of all, to determine the
mechanism generating all the forms of energy densities that are 
detected today.

Starting from the quark-gluon plasma, the evolution of the 
micro and macro worlds can be successfully described in the framework of the 
Standard 
Cosmological Model (SCM) plus Standard Particle Model (SPM). The main features
of such an evolution are in a very good agreement with laboratory and 
astrophysical observations. Nevertheless, in this framework there are no 
answers to the questions of where the primordial plasma came from and how were 
seeded such large structures like voids, galaxies, clusters and 
superclusters of galaxies. According to a top-bottom scheme, these 
structures are
necessary to explain the formation of stars, planets, heavy chemical elements 
and life. To find the missing answers it seems to be unavoidable to go beyond 
the standard paradigms.  

The analysis of recent results of the cosmic microwave background (CMB) 
observations \cite{CMBdata} confirm the inflationary
scenario as the leading 
candidate for a theory of the formation of the primordial plasma and, 
through gravitational instability, of large-scale structures in our 
Universe. 
According to this theory, an special kind of matter with a characteristic 
negative pressure produced a period of rapidly accelerated
expansion in the early Universe \cite{inflation}. The successive decay of this 
matter, first into non-relativistic particles and then into hot radiation, 
together with the breaking of fundamental symmetries led to the period known
as Hot Big Bang.  
The fluctuations of matter and space-time 
 taking place in the very early Universe \cite{perturbations} were stretched 
out of
the causal horizon by the accelerated expansion and, after reentering the 
Universe in a later era, give 
rise to curvature perturbations, seeding the formation of large-scale 
structure and imprinting the anisotropies in the CMB radiation. 
The simplest choice for the inflationary matter is a single real scalar field, 
coined inflaton, with dynamics dominated by the potential energy. 

The primordial fluctuations are characterized in terms of 
power spectra of scalar and tensor perturbations. Inflaton driven expansion 
does not give rise to vectorial perturbations. Different theories for the 
physics of the very early Universe lead to different primordial spectra 
(different initial conditions for density evolution) and, after evolving, to
 different distributions of
density and temperature anisotropies in the currently observable Universe. 
The level of accuracy of current experiments measuring the CMB is of the order 
of $10\%$,
which roughly corresponds to $10\%$ uncertainty for the best determined 
cosmological parameters. Future experiments \cite{future} will increase 
this accuracy to the limit of the cosmic variance what must allow to strengthen
 the constrains upon the cosmological parameters, including the inflationary 
ones.
This way, the number of suitable inflationary potentials could be reduced 
to a few candidates, and it could be possible
to obtain some hints about physics on energies close to the Planck scale
if, as we assume throughout this paper, the single scalar field scenario is
the dominant feature of the very early Universe dynamics.

Several extensions of the SPM are able to reproduce its features 
in a equally satisfactory fashion. Hence, the answer to whether our Universe is
 generic it is related to the probability for a given SPM extension to be the 
actual one. But, even if there exist a deterministic way of 
choosing the appropriate SPM extension, different initial values for the 
inflaton field dynamics will lead to different kind of universes. The initial 
values 
of the scalar field and its derivative with respect to cosmic time determine 
the amount of inflation produced before the decay into radiation as well as 
the form of the perturbations spectra. Hence, they will determine the Universe
topology, its life time, how isotropic and homogeneous it will be, the 
temperature of the primordial plasma and the spectra of particles and 
large-scale structure. It means that, given an inflaton potential 
(more precisely, given a sector of the potential near a vacuum state of the 
theory), 
the fundamental role in the diversification of the cosmological evolution is 
played by the initial conditions. 

In the inflationary scenario the initial conditions for the scalar field 
classical dynamics are set by the mechanism known as eternal inflation 
\cite{eternal}. In most regions of the very early Universe the inflaton rolls
down the potential aiming at a stable configuration. However, there is a finite
probability that in some causally connected regions the quantum fluctuations
 dominate upon the
classical behavior, displacing the scalar field to higher energies values. 
These regions inflate more than those where the inflaton is rolling down. This 
way, large spaces arise which, in turn, can be divided in causally connected 
regions where either 
the classical motion or the quantum fluctuations dominate. In principle, this
process never ends, producing regions eternally inflating into the future and 
other where, 
starting from different initial values, the inflaton decays into hot big bangs,
rising different universes. We shall give more details about the mechanism of
(future) eternal inflation in the next section. 
  
A plausible outcome of Planck and Map satellites measurements could be a 
central constant value for 
the rate between the tensorial and scalar perturbations amplitudes 
\cite{KK} (which, in turn, is related to the tensorial spectral index) as 
well as the possibility of fitting the CMB observations using a 
scale-dependent scalar index \cite{Covi,Hannestad}.
Recently, a model generating perturbations described by a 
scale-invariant tensorial spectral index was introduced \cite{ConstNt}. This 
model 
could be the best fit to near future observations if the scalar index can be 
fairly approximated by a constant at small angular 
scales and departs from that behavior at large scales. Taking into account
that currently a constant scalar index is typically used to fit the CMB 
observations \cite{CMBdata}, these assumptions for its possible 
scale-dependence are well based. Thus, the physics corresponding to this 
potential could be close 
to the actual very early Universe physics. In Ref.~\cite{SchHuat}, assuming a 
strong
similarity between the above mentioned inflationary potential
and the actual one, it was concluded that the universes in the
eternal inflation picture will be very similar each to the other from the
observational point of view.
It means that our Universe will be a generic
outcome of the cosmological evolution rather than some extraordinary event.

Calculations supporting that conclusion rely on several assumptions like
a constant tensorial spectral index, the second order precision of the 
involved expressions, and a semi-classical 
analysis close to the Planck era. In Ref.~\cite{ConstNt} it was shown that
higher order corrections can introduce relevant features in the form of the 
inflationary potential derived from information on the functional forms of
the spectral indices. For example, to leading order it makes no sense to 
look for potentials yielding spectra characterized by a constant tensorial
index and a scale-dependent scalar index, while this solution can be
successfully obtained to next-to-leading order \cite{ConstNt}. Consequently,
it could be expected that corrections beyond the next-to-leading order could 
reveal new important information about the potential form. Nevertheless,  
even if higher order corrections to primordial spectra have been recently 
proposed
for slow-roll models \cite{SG} and some classes of non slow-roll models 
\cite{HFFampl},
the given expressions need to be tested for accuracy before the corresponding
inverse problem can be introduced. So far, this task has not been 
accomplished.
On the other hand, though the most important from the observational point of
view part of inflation occurs at energies substantially lower than Planck
scales, we have seen in Ref.~\cite{ConstNt} that the departure from power-law
behavior takes place in the high energies regime, where the initial conditions
would be set. Thus, one can wonder whether this departure will occurs (and in
 case it occurs, if it will occur in the same fashion) if the corresponding
initial
energy scale is close to the Planck scale. Not having a completely satisfactory
quantum gravity theory, any answer to this question will imply resorting to
additional assumptions. 
With these regards, in this manuscript we would like to focus in the analysis
of the robustness of the conclusion about a 
generic Universe with respect to the
scale-dependence of the tensorial spectral index.
Even regarding near future CMB polarization measurements to be carried out by
Planck satellite \cite{future}, there is not enough motivation for considering 
a scale-dependent tensorial index while analyzing the cosmological data. The 
most optimistic expectations are for
estimating a very small constant central value for this index. From the
theoretical side, it can be shown that the scale dependence of the tensorial
index must be very weak. However, in Ref.~\cite{ConstNt} a weak scale 
dependence for the scalar index was allowed while keeping constant the 
tensorial one, and this lead to a significant difference between the obtained
solution and power-law inflation for high values of the scalar field. This
difference could be a consequence of the high nonlinearity of the involved
equations. Taking this 
into account, the question arises of whether a weak scale 
dependence for the tensorial index could change the inflationary picture in the
high energies regime in such a way that the conclusion about a generic Universe
would be affected.
With this motivation, we introduce in Sec.~\ref{sec:models} an ansatz for the 
tensorial
index consisting of a second order polynomial on the first horizon flow 
function (or first slow-roll parameter) \cite{HFFampl}. With this ansatz we 
solved the 
Stewart-Lyth inverse problem introduced in Ref.~\cite{SLIP}   
and found a couple of models with 
slowly varying spectral indices. One of this models can be regarded as a 
generalization of the scenario described in Ref.~\cite{ConstNt}. In the second 
scenario, the spectral indices do not converge to a constant value but still
have a very weak scale-dependence in a wide range of scales and met all the 
conditions for successful 
inflation. Even if in this scenario different outputs from inflation are 
possible, we argue that the higher probability is for universes expanded
by the same factor than our and with perturbations spectra close to a 
power-law.
In Sec.~\ref{sec:Concl} we summarize our discussion. 

\section{Future-eternal inflation}
\label{sec:eternal}

An important effect of the rapidly accelerated primeval expansion is the 
almost instantaneous flattening of the Universe. Hence,
during inflation it can be assumed the space-time geometry 
being described by flat Friedmann-Robertson-Walker metrics. For a single 
real scalar field the stress-energy tensor is like that of a perfect fluid 
with,
\begin{eqnarray}
\label{eq:T}
\rho&=&\frac{\dot{\phi}^2}{2}+V(\phi) \, , \\
p&=&\frac{\dot{\phi}^2}{2}-V(\phi) \, , 
\end{eqnarray}
where $\rho$ is the energy density, $p$ is
the pressure, $\phi$ is the inflaton, and $V(\phi)$ the inflationary
potential. Dot stands for derivatives with respect to cosmic
time. Then, looking for a negative pressure we must have $V>\dot{\phi}^2$.
Combining the temporal and spatial Einstein equations is obtained,
\begin{equation}
\label{eq:Friedmann}
H^2 = \frac{\kappa}{3}\left(\frac{\dot{\phi}^2}{2} + V(\phi)\right),
\end{equation}
while the energy conservation law, yields,
\begin{equation}
\label{eq:mass}
\ddot{\phi} + 3H\dot{\phi} = -V^\prime(\phi) \, .
\end{equation}
Here, $H=\dot{a}/a$ is the Hubble parameter, $a$ the scale
factor, prime stands for derivatives with respect to $\phi$, 
$\kappa = 8\pi/m_{\rm Pl}^2$ is the
Einstein constant and $m_{\rm Pl}$ the Planck mass. Note that for a unique
solution of this system it is necessary to set the initial conditions as for
$\phi$ as for $\dot{\phi}$.

Since during inflation the coefficient in the friction term of 
Eq.~(\ref{eq:mass}) is almost constant then, for
the inflationary period to be long enough, the potential must be rather flat.
Inflation ends when the inflaton reaches the true vacuum which, without loss of
generality, we shall assume located at the origin. An alternative
mechanism for finishing inflation involves several secondary scalar fields 
frozen in unstable states during the expansion until the inflaton reaches a 
critical value which triggers the decay of the secondary fields \cite{hybrid}. 
Thus, inflation must start far enough from the value corresponding to the 
potential absolute minimum (or from the triggering critical value). 
That
condition imposes some tunning of the initial conditions that it is the kind 
of problem the inflationary idea is aimed to solve.
This paradox is resolved assuming a random initial spatial distribution
for the scalar field values. Due to a phenomenon known as (future) eternal 
inflation
\cite{eternal} which occurs in the early era of the expansion,    
it is sufficient with a tiny 
patch of the very early Universe having the required inflaton value. Being the 
inflaton a quantum object, in a region of linear size $d_H=H^{-1}$ (a 
region which may be in causal contact within one Hubble time, 
$\Delta t = H^{-1}$, and we shall call it a Hubble region), the time 
evolution of the average value of $\phi$  can be described by,
\begin{equation}
\phi(t + \Delta t) = \phi(t) + \Delta_{cl}\phi(\Delta t) 
+ \delta_{qu}\phi(\Delta t) \, ,
\label{InfDyn}
\end{equation} 
where $\phi(t+\Delta t)$ and $\phi(t)$ are the inflaton values 
at times $t+\Delta t$ and $t$, 
$\Delta_{cl}\phi(\Delta t)\simeq \dot{\phi}H^{-1}$ 
and 
$\delta_{qu}\phi(\Delta t)$ are changes of the inflaton value during a Hubble 
time due to the classical motion and quantum fluctuations
respectively. The random quantum jump $\delta_{qu}\phi(\Delta t)$ is 
characterized by a Gaussian distribution with zero mean and standard deviation 
$\Delta_{qu}\phi\simeq (2\pi)^{-1}H$. Taking into account that 
$V>\dot{\phi}^2$, then, to leading order, $V\simeq H^2/\kappa$ and
larger $\left|\phi\right|$ values will allow larger deviations of $\phi$ from 
the 
average value. If we assume that in some Hubble regions the inflaton can
pick an initial value $\phi_i$ such that, 
\begin{equation}
\label{eq:Cond4Roll}
{\Delta_{qu}\phi(\phi_i) \over \Delta_{cl}\phi(\phi_i)} \simeq
\frac1{2\pi}\frac{H^2}{\dot{\phi}} = 
\frac1{2\pi}\frac{H}{\frac{d\phi}{dN}}\geq 1 \, ,
\end{equation}
then we have a finite
probability (of about a 16\%) for the negative (positive) quantum contribution 
to
be larger than the positive (negative) classical displacement. In 
Eq.~(\ref{eq:Cond4Roll}) $N \equiv \ln (a/a_{\rm i})$ is the number of efolds 
by which the size of the patch is increased since some initial time 
$t_{\rm i}$. (Note that usually the efolds number is counted backward in time, 
we count it forward, i.e., $N(t_{\rm i})=0$.) The conclusion
is that there is a finite 
probability for the existence of some Hubble regions in the very 
early Universe where the overall scalar field motion will be in the direction 
of larger 
$\left|\phi\right|$ values, i.e., higher energies. During inflation, larger
values of $\left|\phi\right|$ correspond to a larger factor
 $e^N$. 
Then, regions 
where the scalar field rolls upward inflate more and after
some Hubble times we will have large spaces which can be divided again in 
Hubble regions, some of them with a finite probability for the inflaton to
diffuse up the potential, and so on.    
In regions
where the overall scalar field motion is in the direction of smaller 
$\left|\phi\right|$ values, i.e., lower energies, the inflaton eventually will
 reach the potential minimum (or the triggering critical value) and will give
 rise 
to big bangs leading to different universes. Thus, the general picture 
involves some regions eternally inflating and a 
large number of universes arising through big bangs, with their 
large-scale structures seeded by primordial curvature perturbations and 
growing due to gravitational instability. 

For each one of these universes the inflaton final roll-down 
starts with different values of $\phi$ and $\dot{\phi}$ and, according with 
Eq.~(\ref{eq:mass}), it means that the inflationary period will set different
initial conditions for the big bang processes. Notice, however, that the 
mechanism of future eternal inflation provides that most of the initial 
conditions are set at the higher absolute values of the scalar field. Even if 
we assume that the whole set of initial conditions leads to a unique
vacuum state, after inflation
there will be
universes with different degree of flatness, and/or different degree of 
homogeneity and isotropy, and/or different rate of expansion, and/or different
 number of
topological defects, and/or different number and spectrum of particles, and/or
different
large scale-distribution of matter. Summarizing, from the observational point 
of view, most of these 
universes are different each from the other. Therefore, the question is which 
is the 
probability for a universe like our to exist in this multiverses scenario. It
can be naively concluded that with an infinite number of universes
this probability must be high, but it is not so simple. The point is that, in
the definition
\[
Probability = {number \; of \; universes \; like \; our \over
total \; number \; of \; universes} \, ,
\]
we are dealing with infinite quantities, $\infty/\infty$, i.e., it is an 
ill-defined probability.

\section{Weakly scale-dependent spectral indices}
\label{sec:models}

In order to assess the influence that different initial conditions around a
given vacuum state exert upon 
the results of the inflationary era it
is necessary to know which is the inflationary potential. However, this is 
perhaps
the ultimate quest in the very early Universe research. No particle physics 
based theory has 
been completely successful in explaining inflation \cite{PartPhysInf}. 
An alternative to solve the problem we are concerned about it is to look for
common features of those potentials likely to fit the results of
CMB measurements.
In Ref.~\cite{SLIP} a procedure coined Stewart-Lyth inverse problem (SLIP) was 
introduced to determine the potential corresponding to given functional 
forms for the spectral indices. 
In the current analysis of observations, a constant scalar spectral index close
 to unity and zero primordial 
gravitational waves contribution have been used to fit the CMB spectrum with a
rather remarkable success \cite{CMBdata}. It implies that in the inflaton 
range corresponding to the horizon-crossing out of scales currently observed,
 the potential must look like a very flat exponential typical of power-law 
inflation \cite{PLinfl}. Obviously,
 a large 
number of functional expressions for the potential can have the appropriate 
form.  

On the other hand, from upcoming measurements of CMB polarization to be done by
 satellite Planck it
could be possible to determine a central constant value for the rate between
tensorial and scalar inflationary amplitudes. From this value it is 
possible to estimate a constant leading order value for the tensorial index 
\cite{KK}. Furthermore, it was already found 
currently available data to be compatible with scale-dependent 
scalar index \cite{Covi}, though this dependence seems to be highly constrained
\cite{Hannestad}. The increased range and resolution of the CMB observations 
must allow to discern the scale-dependence of the scalar index. If the 
leading order of tensor-scalar ratio is measured and the scale-dependence of
the scalar index is hinted, then stronger constraints in the potential 
functional form will be set. 

The SLIP procedure was used to find the 
inflationary scenario, other than power-law, with constant 
tensorial spectral index \cite{ConstNt}. The 
corresponding scalar index was found to be almost constant at small angular
scales but diverges from that behavior at large angular scales. Therefore, 
the 
obtained potential could be a good approximation to the actual inflaton 
potential. The 
theoretical potential is given as a parametric function described by two 
branches. In Ref.~\cite{SchHuat} it was shown that the form of these branches 
can be fitted by
some hybrid inflation potentials with running exponential couplings (see report
\cite{PartPhysInf} for suitable patterns and related physics). If the
similarity between the parametric potential and the actual one is realistic, 
then it
can be shown that different initial conditions do not lead to different 
outcomes from the inflationary era. Therefore, our universe would
be a generic result of the cosmological evolution rather than an accident 
\cite{SchHuat}. Nevertheless, the assumption of a constant tensorial 
index is justified by measurements limitation but not by theoretical judgment.
From the theoretical point of view some scale-dependence for this index must 
be expected, even if it is very weak to be observed. Moreover, the nonlinearity
of the SLIP related equations may amplify any differences in the
input of the inverse problem. That amplification could
happen in such a way that the
SLIP solutions corresponding to closely related functional forms of the 
spectral indices could be very different each from other. It means that a 
conclusion
drawn from the analysis of one SLIP solution may not be true for other 
SLIP solution, even if the input of the corresponding inverse problems are 
slightly different. Taking this into account, is of interest to 
test the robustness of the prediction of a generic Universe drawn in 
Ref.\cite{SchHuat} by allowing a weak scale dependence of the tensorial index.

Constant spectral indices correspond to a power-law scenario of inflation 
where they are given by \cite{PLinfl,ConstNt},
\begin{equation}
\delta = \Delta = \frac{\epsilon_1}{\epsilon_1-1}=
-\left(\epsilon_1+\epsilon_1^2+\epsilon_1^3+\cdots\right) \leq 0
\, ,
\label{eq:PLind}
\end{equation}   
where $\delta=n_T/2$, $\Delta=(n_S-1)/2$, $n_T$ and $n_S$ are the tensorial and
scalar spectral indices respectively, and the first horizon flow function 
\cite{HFFampl}
\begin{equation}
\label{eq:e1}
\epsilon_1 \equiv {{\rm d} \ln d_{\rm H}\over {\rm d} N} =
 3 { \frac{{\dot{\phi}}^2}{2} \over \frac{{\dot{\phi}}^2}{2} + V(\phi)}
\, ,
\end{equation} 
measures, in general, the logarithmic change of the Hubble distance per e-fold 
of expansion and, particularly in the case of inflaton driven expansion, the
 contribution of the scalar field kinetic energy to the total field energy. 
Inflation proceeds for $\epsilon_1 < 1$ 
(equivalent to $p<0$ and $\ddot{a} > 0$) and, on the other hand, 
$\epsilon_1 >0$ from the weak 
energy condition (for a spatially flat universe). For power-law inflation, 
$\epsilon_1=\rm const.$.

With this regard, we propose the following ansatz for the tensorial index,
\begin{equation}
\label{eq:dansatz}
\delta(\epsilon_1)=-\left[(1+a)\epsilon_1^2+(1+b)\epsilon_1+c\right]
\, ,
\end{equation}
where $\epsilon_1$ is now a scale-dependent function and $a,b,c$ are real 
numbers. The second order of $\delta$ is for compatibility 
with the order of the algebraic equations for the spectral indices
\cite{SL} from which the SLIP differential equations are 
derived. Taking into account that $\delta(\epsilon_1)<0$ \cite{ConstNt}, the
extremum of the parabola given by Eq.~(\ref{eq:dansatz}) must be a maximum.
Combining this with condition $\epsilon_1>0$ it is obtained that $1+a>0$ 
and $b+1<0$. On the other hand, combined with condition $\epsilon_1<1$ it give
us that $1+b>-2-2a$. Finally, requiring $\delta\leq 0$ while evaluated at this 
maximum yields $c\geq (1+b)^2/4(1+a)\geq 0$. Summarizing we have, 
\begin{equation}
\label{eq:abc}
a>-1 \, , -1>b>-3-2a \,\,\,({\rm or}\,\,\, b\geq -2\sqrt{(1+a)c}-1) 
\, , c>0 \, .
\end{equation}
We shall see that two different cases arise corresponding to $b^2>4ac$ and
$b^2<4ac$.  

\subsection{Case 1, $b^2>4ac$}

We start by analyzing case $b^2>4ac$. As we shall see, the result for this case
is close to that obtained in Ref.~\cite{ConstNt} for
a constant tensorial index. In fact, $0\geq a>-1$ is just a 
generalization of that SLIP solution. Thus, the conclusion about a generic 
Universe drawn in Ref.~\cite{SchHuat} can be trivially extended to this model
when $0\geq a>-1$. With this regard, we shall focus
here in the SLIP solution for $a>0$, where the corresponding expression for the
potential will be slightly different.
For $a>0$ the strongest condition on $b$ is $b\geq-2\sqrt{(1+a)c}-1$.
Values of $a,b,c$ can be additionally constrained if more 
conditions for successful inflation are taken into account. Firstly, as it will
be shown below, the values
of these coefficients determine the central values for $n_T$ and $n_S$. A range
 for the last can be estimated from the CMB spectrum and averaging the results
 in Ref.~\cite{CMBdata} we obtain $n_S\in[0.87,1.07]$. 
 Without loss of generality
in our analysis, for all of the plots and numerical 
estimates in this subsection, we have chosen $a=0.35$, $b=-1.1$ and $c=0.01$.

\begin{figure}[ht]
\centerline{\psfig{file=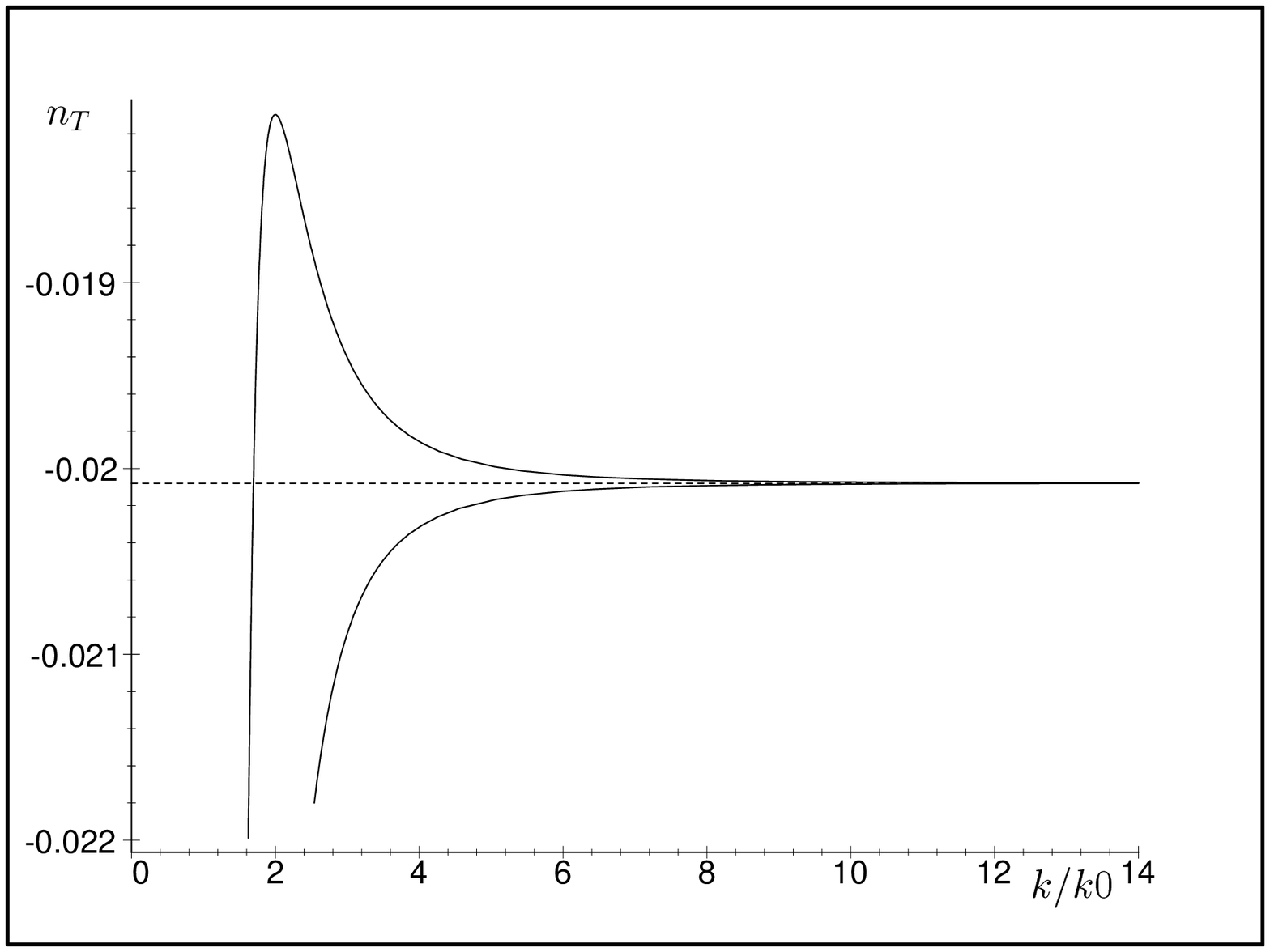,width=8.5cm}
\psfig{file=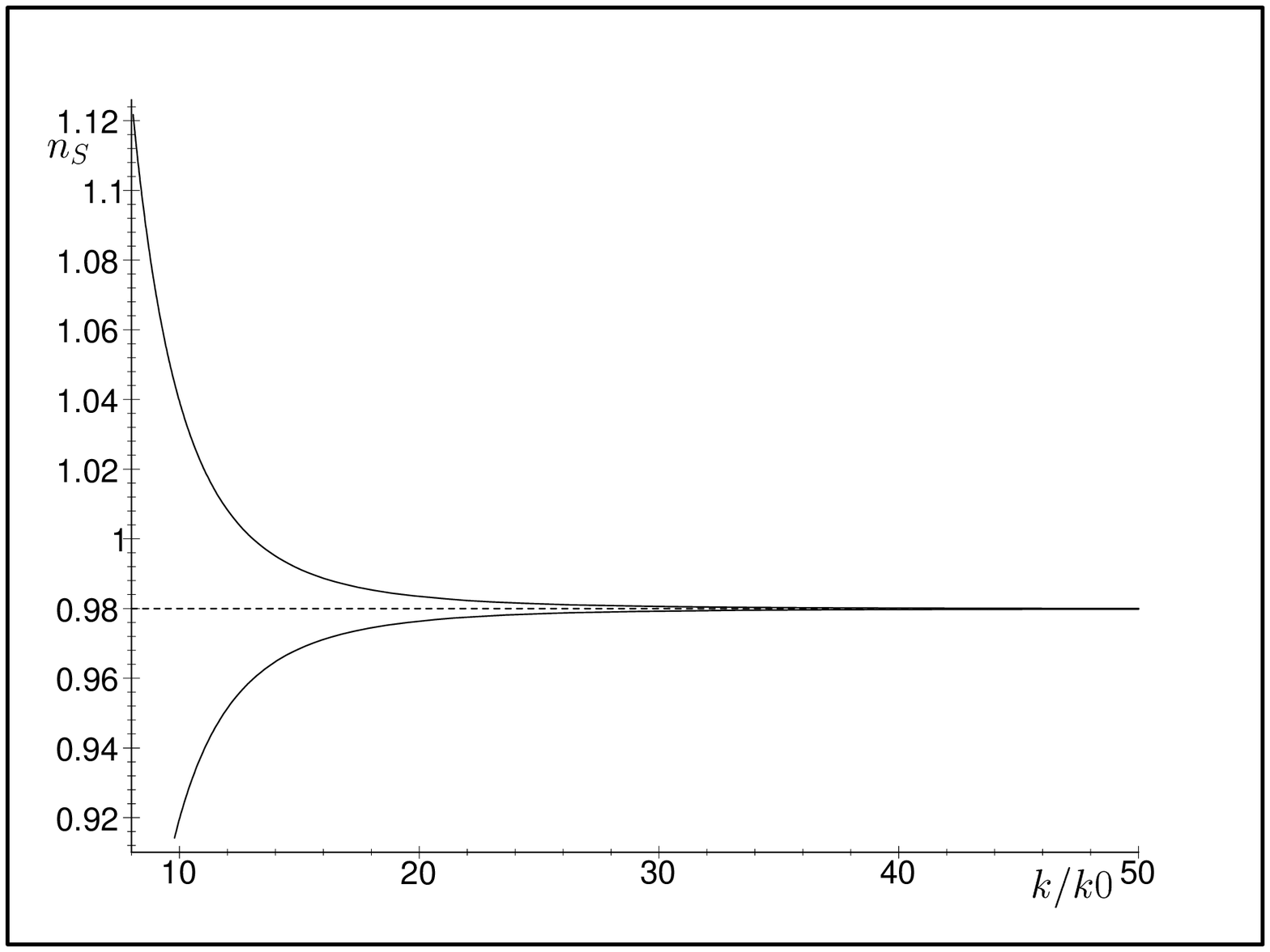,width=8.5cm}}
\caption{Case 1. Plots of the tensorial index (left) and scalar index (right) 
as functions of the horizon crossing wavenumber.}
\label{fig:C1dDk}
\end{figure}

\begin{figure}[ht]
\centerline{\psfig{file=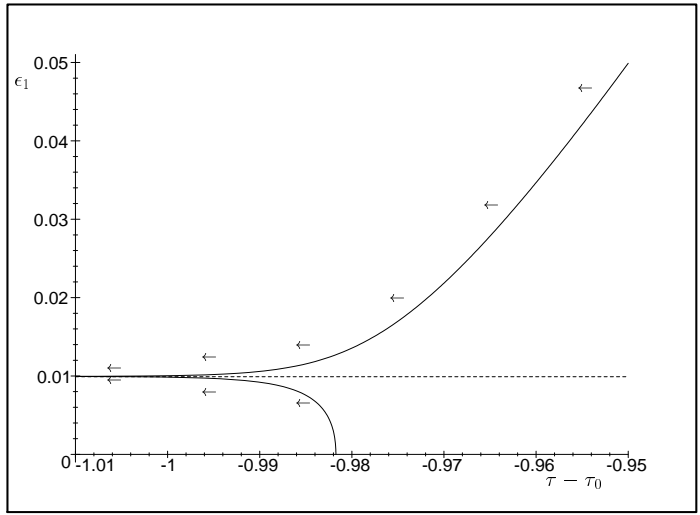,width=9.5cm}}
\caption{Case 1. Plot of $\epsilon_1$ as function of $\tau$. Time flow along
each branch is represented by the small arrows.}
\label{fig:C1et}
\end{figure}

To determine $n_S(\epsilon_1)$ we
use the SLIP equations \cite{SLIP},
\begin{eqnarray}
\label{eq:MSch1}
2C\epsilon_1 \hat{\hat{\epsilon_1}}-(2C+3)\epsilon_1 \hat{\epsilon_1}
-\hat{\epsilon_1}
+\epsilon_1^{2}+\epsilon_1 +\Delta &=&0\,,  \\
\label{eq:MSch2}
2(C+1)\epsilon_1\hat{\epsilon_1}-\epsilon_1^{2}-\epsilon_1 -\delta &=&0\,,
\end{eqnarray}
where $C=-0.7296$ and a hat denotes derivatives with respect to 
$\tau\equiv\ln H^2$. Substituting the ansatz (\ref{eq:dansatz}) in 
Eq.~(\ref{eq:MSch2}) we can separate $\hat{\epsilon_1}$. Deriving with 
respect to $\tau$ we obtain an expression for $\hat{\hat{\epsilon_1}}$. Then,
$\hat{\epsilon_1}$ and $\hat{\hat{\epsilon_1}}$ can be substituted in 
Eq.~(\ref{eq:MSch1}) and
$\Delta$ as function of $\epsilon_1$ will be given by,
\begin{eqnarray}
\label{eq:D}
\Delta(\epsilon_1)&=& -\frac 1{2(C+1)^2}\left\{
\left[Ca^2+\left(2C+3\right)\left(C+1\right)a+2(C+1)^2\right]\epsilon_1^2 
\right.
\nonumber \\
&+&
\left[\left(C+1\right)a+Cab+\left(2C+3\right)\left(C+1\right)b+2(C+1)^2\right]
\epsilon_1
\nonumber \\
&+&
\left. \left(C+1\right)b + \left(2C+3\right)\left(C+1\right)c
+ \left(C+1\right)c\frac 1\epsilon_1
- Cc^2\frac 1\epsilon_1^2 \right\} 
\, .
\end{eqnarray}
On the other hand, 
the wavenumber $k$ at horizon crossing, i.e., $k=aH$, as function of
$\epsilon_1$ can be obtained as solution of equation,
\begin{equation}
\label{eq:ediffk}
(C+1)(\epsilon_1-1)\tilde{\epsilon_1}-\epsilon_1^2-\epsilon_1-\delta=0,
\end{equation}
where $\tilde{\epsilon_1} \equiv d\epsilon_1/d\ln k$ \cite{SLIPk}. Using ansatz
(\ref{eq:dansatz}) and after integration,
\begin{equation}
\label{eq:C1ke}
k=k_0\left|a\epsilon_1^2+b\epsilon_1+c\right|^{-\frac{C+1}{2a}}
\left|\frac{2a\epsilon_1+\sqrt{b^2-4ac}}{2a\epsilon_1-\sqrt{b^2-4ac}}\right|
^{-\frac{(C+1)(2a+b)}{2a\sqrt{b^2-4ac}}} \, ,
\end{equation}
where $k_0$ is the integration constant. Hereafter, the integration constants
will be denoted by the subscript $0$ in the corresponding variable.
The plots of $n_T(k)$ and $n_S(k)$, using $\epsilon_1$ as parameter, are 
presented in Fig.~\ref{fig:C1dDk} where the scale range was chosen to allow
details observation at large scales (small $k$) but it can be straightforwardly
extended to infinity.
As it can be observed, at small angular scales the indices behave roughly as
constants. Two branches can be noted at large angular scales departing from the
 power-law like behavior. To understand the origin of these branches we must
analyze the dynamics of $\epsilon_1$. The solution of Eq.~(\ref{eq:MSch2}) with
$\delta$ given by Eq.~(\ref{eq:dansatz}) is
\begin{equation}
\label{eq:C1te}
\tau=\ln\left[\left|a\epsilon_1^2+b\epsilon_1+c\right|^{-\frac{C+1}{a}}
\left|\frac{2a\epsilon_1+b+\sqrt{b^2-4ac}}
{2a\epsilon_1+b-\sqrt{b^2-4ac}}\right|
^{-\frac{(C+1)b}{a\sqrt{b^2-4ac}}}\right] + \tau_0 \, .
\end{equation}
This solution diverges when 
$\epsilon_1\rightarrow\epsilon_1^\pm$ where, $\epsilon_1^\pm$ are the roots of
$a\epsilon_1^2+b\epsilon_1+c$. Both these roots are positive for $a>0$, $b<0$ 
and $c>0$ but 
for most suitable values of $a$, $b$ and $c$ only $\epsilon_1^-$ will be less 
than $1$. In Fig.~\ref{fig:C1et}, 
$\epsilon_1(\tau)$ as given by solution (\ref{eq:C1te}) is plotted.
We observe that the branches in Fig.~\ref{fig:C1dDk} 
correspond to initial values for $\epsilon_1$ which are greater or less than
$\epsilon_1^-$. This value is shown in all the figures of this subsection by 
mean of a dashed line. Recalling that $d\tau/dt<0$, independently of the 
initial conditions, with cosmic time the solutions asymptotically converge to 
the solution given by $\epsilon_1=\epsilon_1^-$. This can be observed in 
Fig.~\ref{fig:C1et} 
where the time flow is represented by small arrows near the
corresponding branch (also used in the remaining figures of this subsection). 
Substituting the
value for $\epsilon_1^-$ in Eqs.~(\ref{eq:dansatz}) and (\ref{eq:D}) it is
obtained that,
\begin{equation}
\label{eq:PLasymp} 
\Delta(\epsilon_1^-)=\delta(\epsilon_1^-)= 
\frac c a + \left( \frac b a - 1 \right) \epsilon_1^-
\end{equation}
confirming that the solution converges to a power-law solution. This asymptotic
behavior is consistent with the result by Hoffmann and Turner \cite{HT} of
power-law being an attractor of the inflationary dynamics. However, we shall
show in the next subsection that it is not generally true.
Expression (\ref{eq:PLasymp}) can be used to fit the values of $a$, $b$ and 
$c$ for $\Delta$ to match the power-law value required to fit the CMB spectrum.

\begin{figure}[ht]
\centerline{\psfig{file=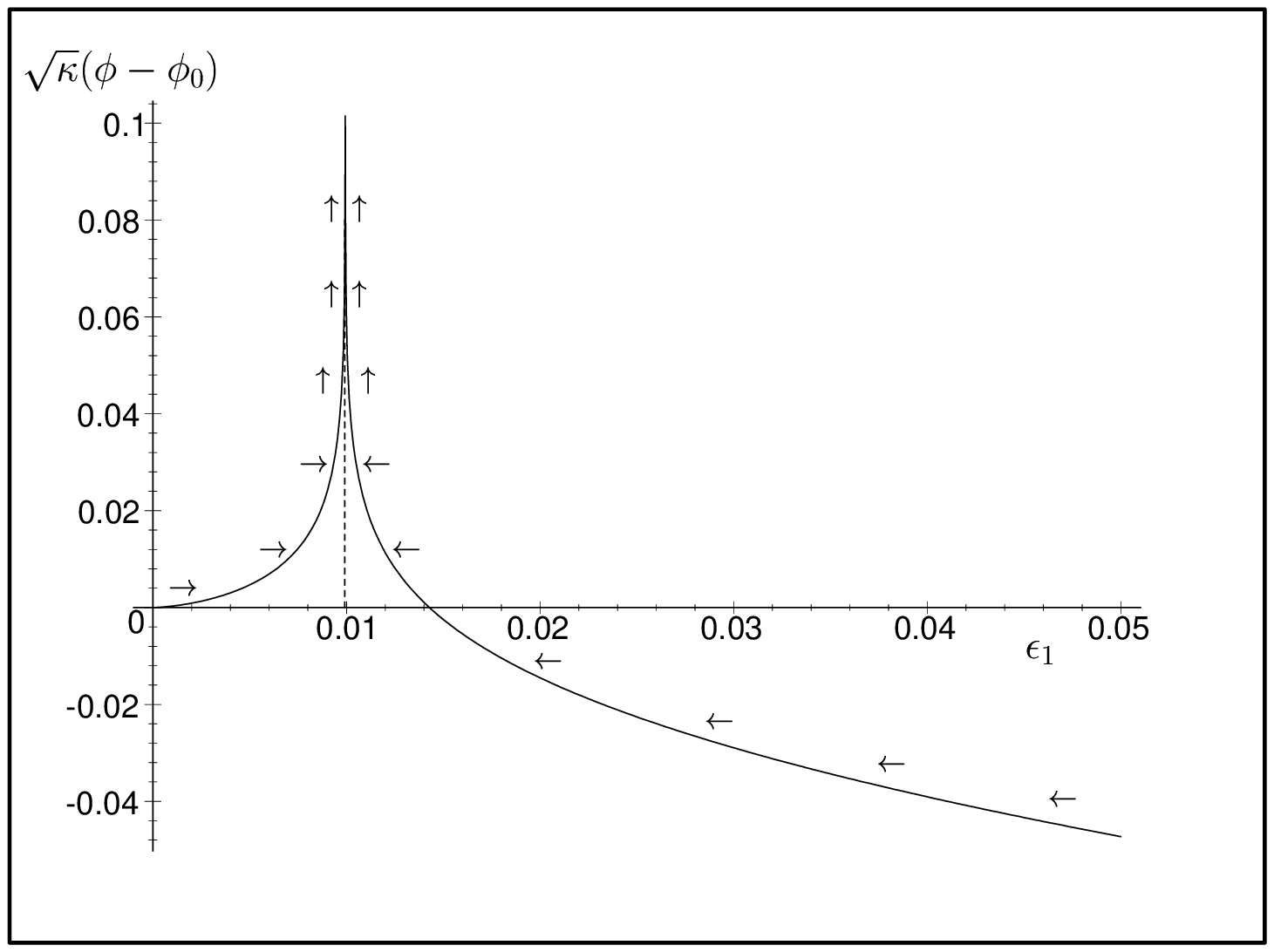,width=8.5cm}
\psfig{file=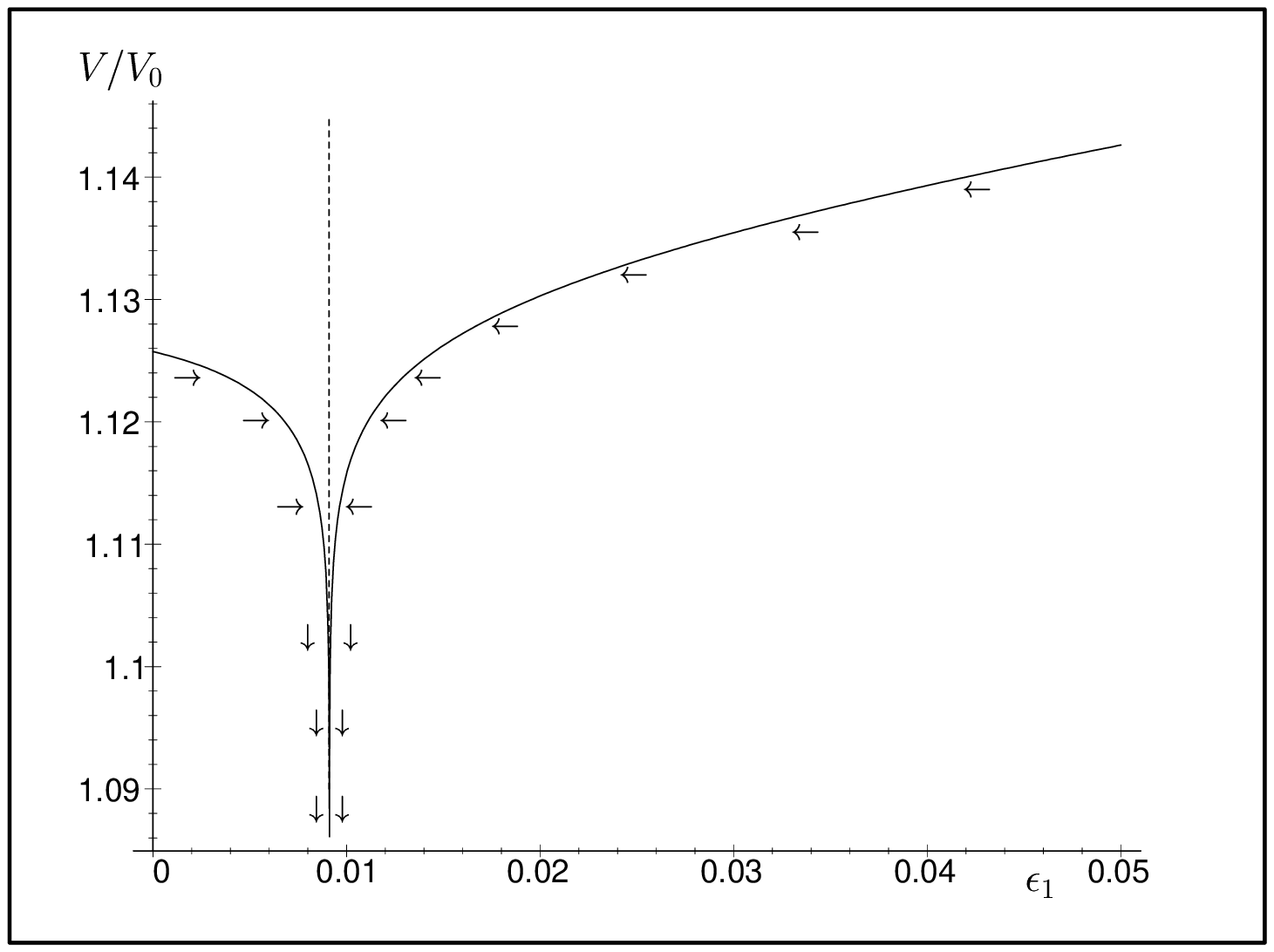,width=8.5cm}}
\caption{Case 1. Plots of the inflaton (left) and the potential (right) as 
functions of $\epsilon_1$.}
\label{fig:C1pVe}
\end{figure}

\begin{figure}[ht]
\centerline{\psfig{file=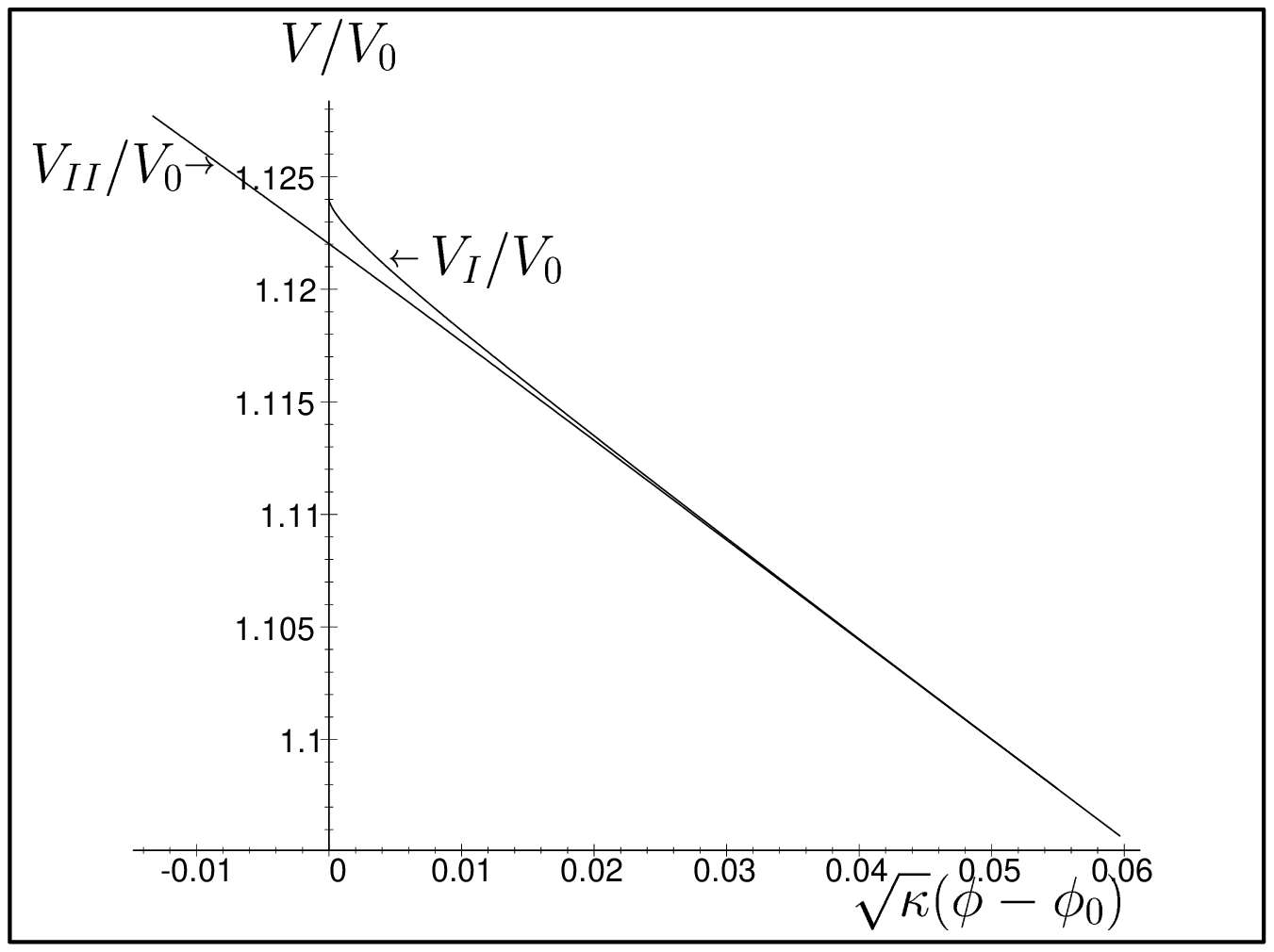,width=9.2cm}
\psfig{file=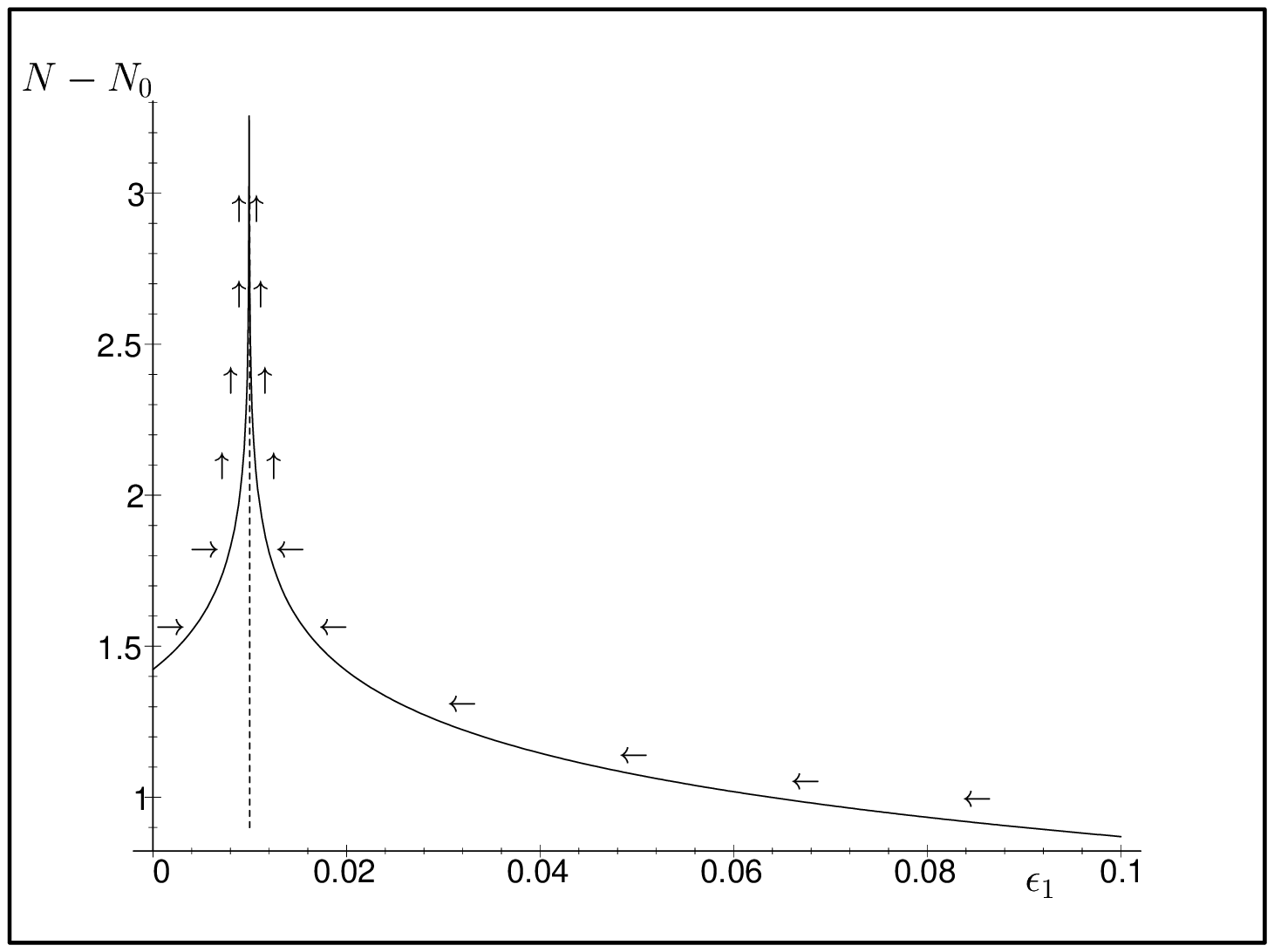,width=8.5cm}}
\caption{Case 1. Plots of the potential as function of $\phi$ (left) and the
efolds number as function of $\epsilon_1$ (right).}
\label{fig:C1VpDNe}
\end{figure}

Recall that as functions of $\epsilon_1$, the scalar field and its potential
are given by \cite{ConstNt},
\begin{eqnarray}
\label{eq:PotentialE}
V(\epsilon_1)&=& \frac{1}{\kappa}\left(3-\epsilon_1\right)
\exp\left[\tau(\epsilon_1)\right]
\, ,
\\
\label{eq:Phi}
\phi(\epsilon_1)&=& -\frac{2(C+1)}{\sqrt{2\kappa}}
\int\frac{\sqrt{\epsilon_1}d\epsilon_1}{\epsilon_1^2+\epsilon_1+\delta}
+ \phi_0\, .
\end{eqnarray}
Thus, for our case, the parametric potential is
\begin{equation}
V(\phi)= \left\{
\begin{array}{rcl}
\phi(\epsilon_1)&=&\frac{(C+1)}{\sqrt{\kappa}}\frac{\sqrt{a}}{\sqrt{b^2-4ac}}
\left[-\sqrt{-b+\sqrt{b^2-4ac}}
\ln\left|\frac{\sqrt{2a\epsilon_1}+\sqrt{-b+\sqrt{b^2-4ac}}}
{\sqrt{2a\epsilon_1}-\sqrt{-b+\sqrt{b^2-4ac}}}\right|\right.\\
&& + \left. \sqrt{-b-\sqrt{b^2-4ac}}
\ln\left|\frac{\sqrt{2a\epsilon_1}+\sqrt{-b-\sqrt{b^2-4ac}}}
{\sqrt{2a\epsilon_1}-\sqrt{-b-\sqrt{b^2-4ac}}}\right|\right]
+ \phi_0 \, ,  \\
&& \\
V(\epsilon_1)&=&V_0(3-\epsilon_1)
\left|a\epsilon_1^{2}+b\epsilon_1 +c\right|^{-\frac{C+1}a}
\left|\frac{2a\epsilon_1+b+\sqrt{b^2-4ac}}
{2a\epsilon+b-\sqrt{b^2-4ac}}\right|^{-\frac{b(C+1)}{a\sqrt{b^2-4ac}}}
\, .
\end{array}
\right.
\label{eq:C1Vp}
\end{equation} 
Plotting $\phi(\epsilon_1)$
and $V(\epsilon_1)$ as in Fig.~\ref{fig:C1pVe}
and comparing with Fig.~\ref{fig:C1et}, one can see that the necessary 
criterion \cite{ConstNt},
\begin{equation}
\left\{
\begin{array}{rcl}
\hat{\epsilon}\frac{d\phi}{d\epsilon}&<&0\, ,  \\
\\
\hat{\epsilon}\frac{dV}{d\epsilon}&>&0\, ,
\end{array}
\right.
\label{eq:epsConds}
\end{equation}
is fulfilled only in the intervals of $\epsilon_1$,
$I=[0,\epsilon_1^-)$ and $II=(\epsilon_1^-, \epsilon_1^*]$ with 
$\epsilon_1^*$ being
the value where the maximum of $V(\epsilon_1)$ is reached. $I$ and $II$
corresponds to the branches previously mentioned when analyzing the dependence
of $\delta$ and $\Delta$, and can be also observed in the parametric
plot of $V(\phi)$ presented in the left part of Fig.~\ref{fig:C1VpDNe}. We 
shall call these branches respectively as $V_I$ and $V_{II}$.

The asymptotic behavior $\epsilon_1\rightarrow\epsilon_1^-$ should ensure that
the inflaton roll-down along any of these two branches will produce enough
inflation. For inflation to make the 
Universe as flat and homogeneous as our, it is required that $N>60$ 
\cite{inflation}. To see clearly that this is the case, the efolds 
number as function of $\epsilon_1$ can be determined using definition 
(\ref{eq:e1}). Changing variables it is obtained,
\begin{equation}
\label{eq:Ne}
\hat{N}=-\frac1{2\epsilon_1} \, ,
\end{equation}
and after using $\hat{\epsilon_1}$ as given by Eq.~(\ref{eq:MSch2}),
\begin{equation}
\label{eq:DNintde}
N=-(C+1)\int \frac{d\epsilon_1}{\epsilon_1^2+\epsilon_1+\delta} + N_0 \, .
\end{equation}
Note that for a given functional form of the tensorial index it can be 
assessed whether the corresponding inflationary scenario will produce the 
required amount of expansion without previous knowledge on the corresponding 
potential. In our case, the integration of Eq.~(\ref{eq:DNintde}) yields,
\begin{equation}
\label{eq:DNe}
N=-\frac{(C+1)}{\sqrt{b^2-4ac}}
\ln\left|\frac{2a\epsilon_1+b+\sqrt{b^2-4ac}}
{2a\epsilon_1+b-\sqrt{b^2-4ac}}\right| + N_0
\, ,
\end{equation}
with the plot presented in the right part of 
Fig.~\ref{fig:C1VpDNe}. As was expected, the closer $\epsilon_1$ to 
$\epsilon_1^-$, the larger the increase of expansion of the Universe.

This model lacks a graceful exit into the SCM, i.e., there is no
``natural'' way out of inflation here. To solve this problem,  
$\phi$ can be regarded as the dominant scalar field in
a hybrid scenario with $\epsilon_1^-$ being the value corresponding
to the critical value of $\phi$ near which the false
vacuum becomes unstable and the multiple scalar fields roll to the
true potential minimum. It means that for any initial condition with
$\epsilon_1(t_i)$ slightly different from $\epsilon_1^-$ the Universes will
get sufficiently flat and homogeneous and the observable spectra of 
perturbations will be almost of power-law type. 

Let us now see which are the conditions for eternal inflation to take place in
this model. First of all, we note that in general,
\begin{equation}
\frac{d\phi}{dN}=\sqrt{\frac 2 \kappa}\sqrt{\epsilon_1} \, .
\end{equation}
Then, condition (\ref{eq:Cond4Roll}) will read 
\begin{equation}
\label{eq:Cond4EtInf}
\frac{H^2(\epsilon_{1i})}{\pi m_{\rm Pl}^2 \epsilon_{1i}} \geq 1 \, ,
\end{equation}
for some value $\epsilon_{1i}$. For the particular model we are analyzing 
in this subsection, and using $H^2$ from Eq.~(\ref{eq:C1te}), we obtain,
\begin{equation}
\frac{H_0^2}{\pi m_{Pl}^2 \epsilon_{1i}}
\left|a\epsilon_{1i}^2+b\epsilon_{1i}+c\right|^{-\frac{C+1}{a}}
\left|\frac{2a\epsilon_{1i}+b+\sqrt{b^2-4ac}}
{2a\epsilon_{1i}+b-\sqrt{b^2-4ac}}\right|
^{-\frac{(C+1)b}{a\sqrt{b^2-4ac}}} \geq 1\, .
\end{equation}
It can be checked (for example, by plotting the left hand side of this 
relation) that a suitable selection of the scale given by $H_0$ must ensure 
that condition (\ref{eq:Cond4EtInf}) will be met for any value of 
$\epsilon_{1i}$ far enough from $\epsilon_1^-$. $H_0$ must be chosen taking 
into account that the damping of inflaton evolution due to the back reaction of
the particles produced during the expansion must take place far enough from
the quantum dominated regime \cite{DS}.   
Then, for any proper inflationary energy scale $V_0$, the scalar field
starts to roll down the potential in Hubble regions with values of  
$\epsilon_1$ very close to zero or to $\epsilon_1^*$, i.e., regions with larger
values of $|\phi|$. On the other hand,
there is no reason to assume a stochastic initial distribution
for $\phi$ and not for $\dot{\phi}$. According to 
Eq.~(\ref{eq:e1}), starting with random $\phi$ and 
$\dot{\phi}$ means starting with random $\epsilon_1$ then,
in the beginning, in some Hubble regions the inflaton is located in branch 
$V_I$, 
and in some other Hubble regions, in branch $V_{II}$. Therefore, in the subset 
of these regions where the roll-up the potential will eventually dominate, 
different high
energies physics will be set up. Now, while rolling down, $\Delta_{qu}\phi$
decreases and, due to the uncertainty principle, the equal-time standard 
deviation for 
the canonical conjugate of $\phi$, i.e., $\Delta_{qu}\dot{\phi}$ must increase.
Since, for $\epsilon_1\in I$ as well as 
for $\epsilon_1\in II$, at these energies the evolution of $\epsilon_1$ is 
characterized by 
the asymptotic convergence $\epsilon_1\rightarrow\epsilon_1^-$, any quantum 
perturbation of the
canonical momentum $\dot{\phi}$ (translated into quantum perturbation of
$\epsilon_1$) makes possible that the inflaton ``jumps'' from branch $V_I$
to branch $V_{II}$ and vice versa. This way, memories from the 
corresponding
high energies physics will be smoothly erased. Moreover, this asymptotic 
behavior of $\epsilon_1$ ensures that most of the inflaton perturbations were
produced close to the logarithmic end of inflation. These are the perturbations
which play the most important role from the point of view of large-scale
structure formation. As we have already seen, at these scales 
the perturbations spectra produced while rolling down $V_I$ or $V_{II}$ are 
almost identical to power-law spectra.
Therefore, according with potential (\ref{eq:C1Vp}), no matter which the 
initial conditions were, there is a big chance for all the big bang universes 
(including our) being very similar each to the other.

\subsection{Case 2. $b^2<4ac$}
\begin{figure}[ht]
\centerline{\psfig{file=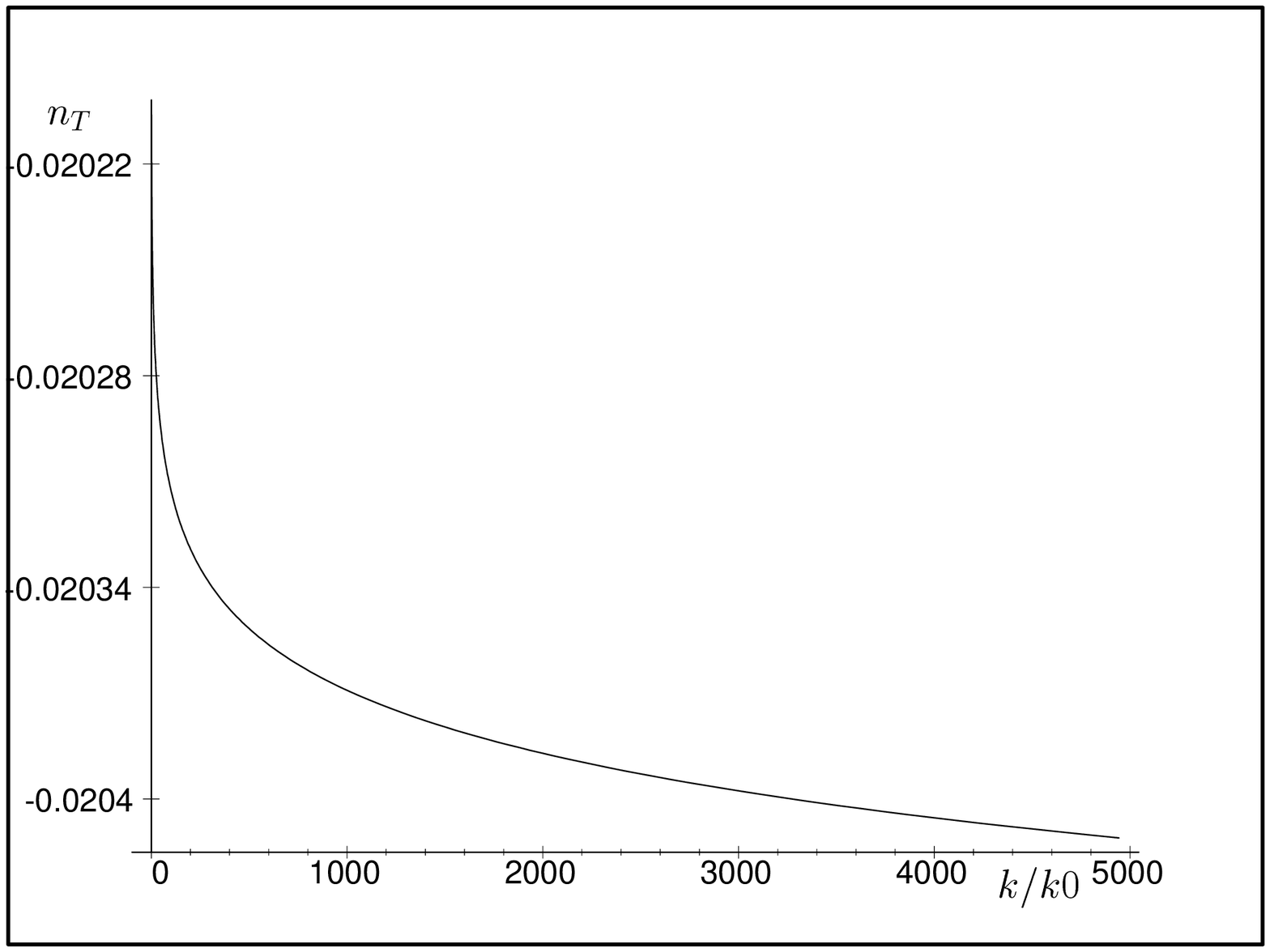,width=8.5cm}
\psfig{file=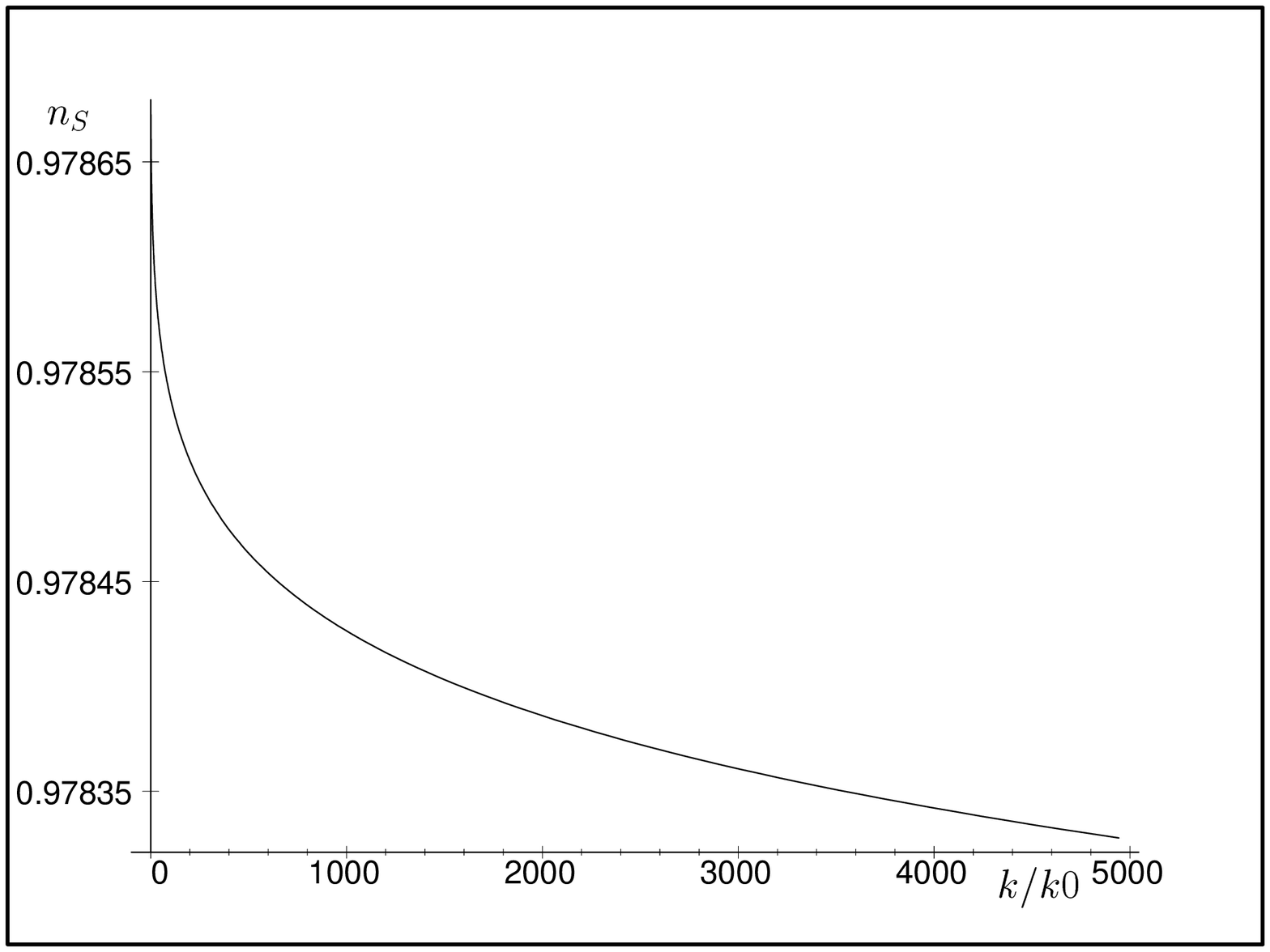,width=8.5cm}}
\caption{Case 2. Plots of the tensorial index (left) and scalar index (right) 
as functions of the horizon crossing wavenumber.}
\label{fig:C2dDk}
\end{figure}

\begin{figure}[ht]
\centerline{\psfig{file=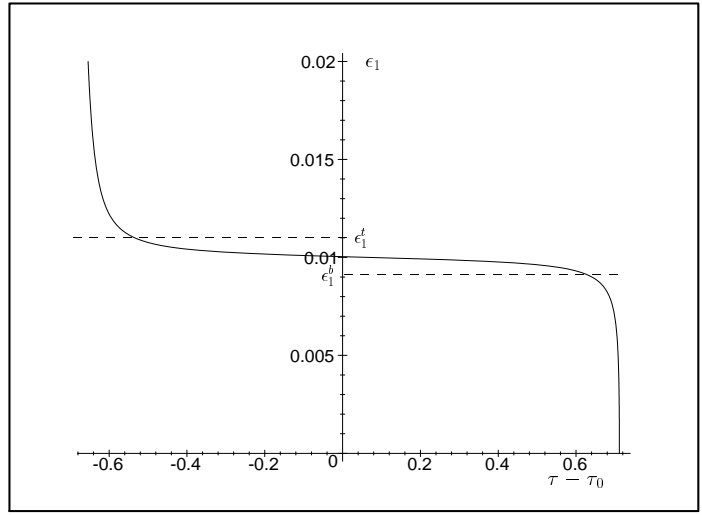,width=9.5cm}}
\caption{Case 2. Plot of $\epsilon_1$ as function of $\tau$.}
\label{fig:C2et}
\end{figure}

In this subsection we shall analyze case $b^2<4ac$ which, as it will be shown,
it is a step further in the relaxation of the scale-dependence of the spectral 
indices while still satisfying the observational bounds for such a dependence.
With respect to the condition labeling this case, new restrictions for
the coefficients values arise, i.e., 
$a>0$, $-1>b>-1/2-2ac$ and $c>1/4a$. On the other hand, we shall see 
that, for a sufficient efolds number we must require $4ac-b^2\ll 1$. Once more,
$c$ will be chosen for $n_S$ to be in the center of $[0.87,1.07]$. In this
subsection the coefficients numerical values are, $a=50.01$, $b=-1.0002$
and $c=0.005004$. For this 
case, the scalar index as function of $\epsilon_1$ is also given by expression
(\ref{eq:D}), but the wavenumber $k$ at the horizon crossing, 
as function of $\epsilon_1$ is now,
\begin{equation}
\label{eq:C2ke}
k=k_0\left|a\epsilon_1^2+b\epsilon_1+c\right|^{-\frac{C+1}{2a}}
\exp
\left[\frac{(C+1)(2a+b)}{a\sqrt{4ac-b^2}}
\arctan\left(\frac{2a\epsilon_1+b}{\sqrt{4ac-b^2}}\right)\right]
\, .
\end{equation}
The corresponding parametric plots of $n_T(k)$ and $n_S(k)$ for a given range
of scales are presented in 
Fig.~\ref{fig:C2dDk}.
It is observed that, within the observational error, both indices can be 
approximated by constant values 
in a wide range of scales but neither of them asymptotically converges to a 
constant. Again, for an understanding of this behavior we must
analyze the dynamics of $\epsilon_1$. Here, the solution of 
Eq.~(\ref{eq:MSch2}) is,
\begin{equation}
\label{eq:C2te}
\tau=\ln\left|a\epsilon_1^2+b\epsilon_1+c\right|^{-\frac{C+1}{a}}+
\frac{2(C+1)b}{a\sqrt{4ac-b^2}}
\arctan\left(\frac{2a\epsilon_1+b}{\sqrt{4ac-b^2}}\right) + \tau_0
\, ,
\end{equation}
and the plot is presented in Fig.~\ref{fig:C2et}.
A period of almost constant $\epsilon_1$ 
($L_2\equiv(\epsilon_1^b,\epsilon_1^t)$), corresponding to near
 power-law inflation, is bounded by periods where $\epsilon_1$ abruptly
increases with cosmic time. This result disagrees with those in Ref.~\cite{HT}.
We believe that this disagreement is explained by the precision of the 
expressions involved in each analysis. The analysis by Hoffmann and Turner was
done using the leading order of the relevant quantities while we are using here
 the next-to-leading order precision. The dynamical analysis presented in 
Ref.~\cite{SLIP} shows that to next-to-leading order the expressions are even 
more nonlinear that to leading order, and this nonlinearity introduces 
further complexity in the inflationary dynamics. As can be observed in the
figures of the reduced phase spaces for the evolution of $\epsilon_1$ presented
Ref.~\cite{SLIP}, for $\Delta<0$ we have a saddle point located in the 
interesting range of values of $\epsilon_1$. It means that an asymptotic 
behavior will be characteristic only of those trajectories very close to the
unstable separatrices (recall that $d\tau/dt<0$). On the other hand, some 
trajectories must be expected that come near the saddle point and then blow up.
This could be the case for the model we are analyzing in this subsection.

\begin{figure}[ht]
\centerline{\psfig{file=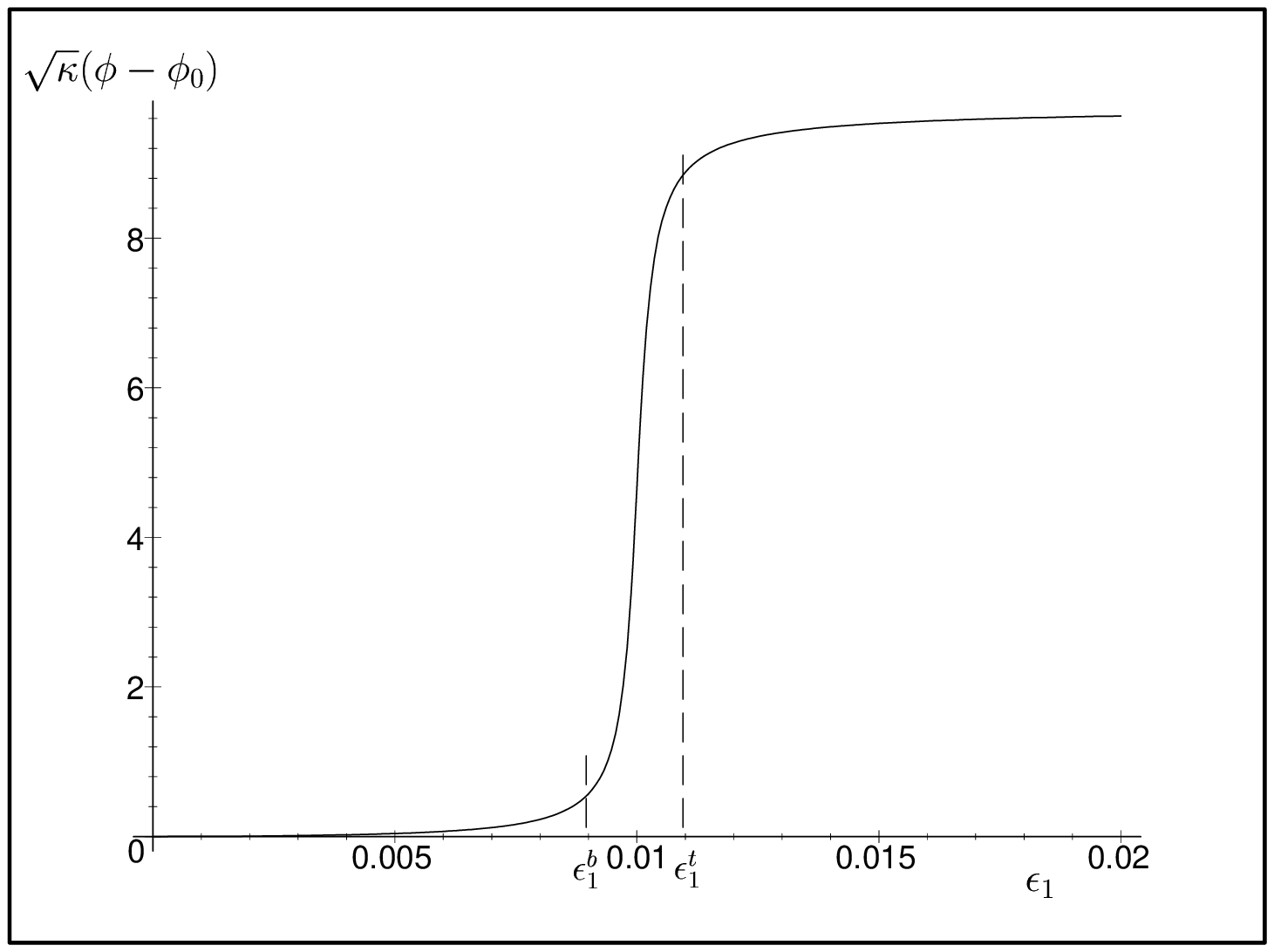,width=8.5cm}
\psfig{file=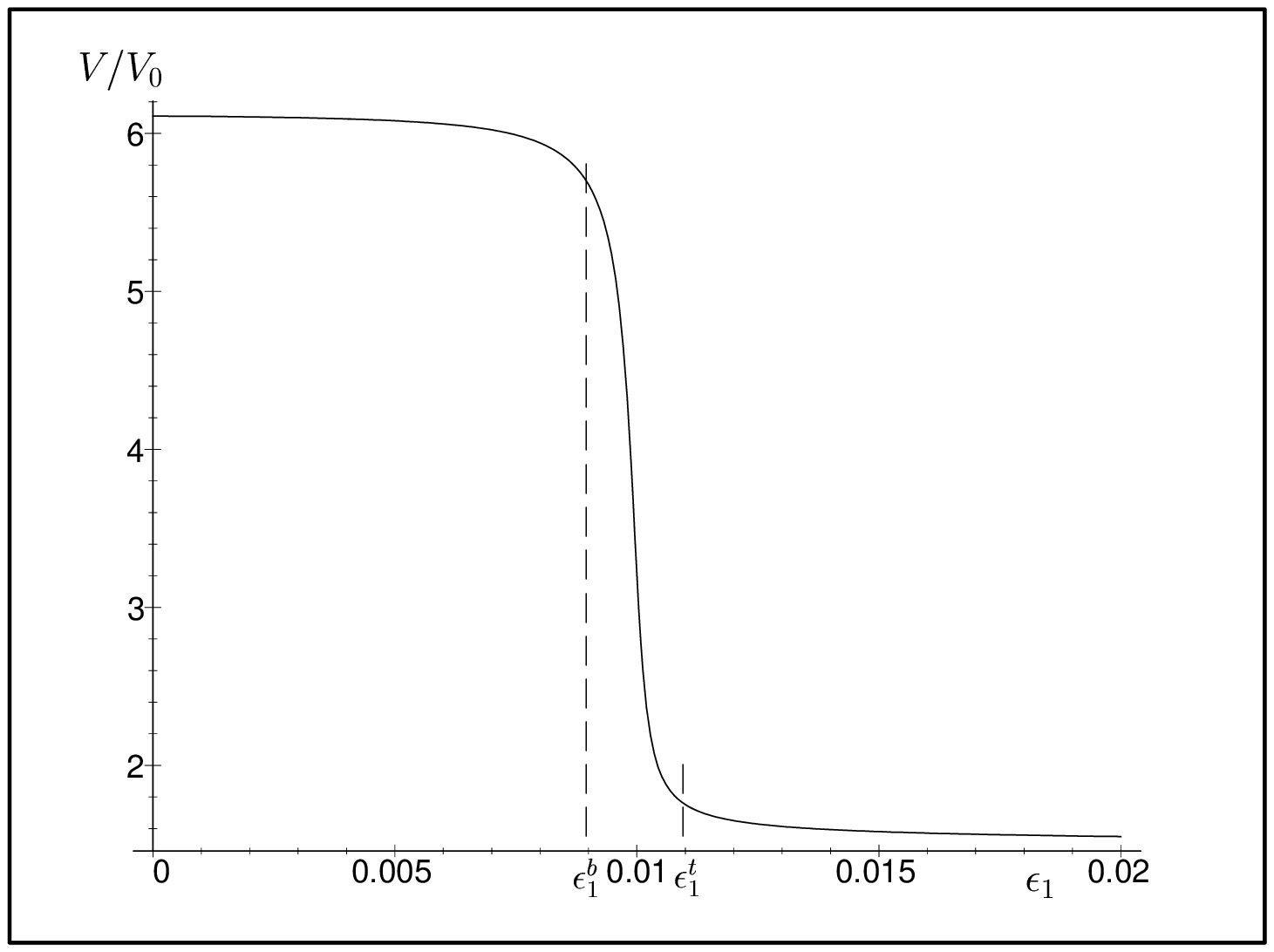,width=8.5cm}}
\caption{Case 2. Plots of the inflaton (left) and the potential (right) as 
function of $\epsilon_1$.}
\label{fig:C2pVe}
\end{figure}

\begin{figure}[ht]
\centerline{\psfig{file=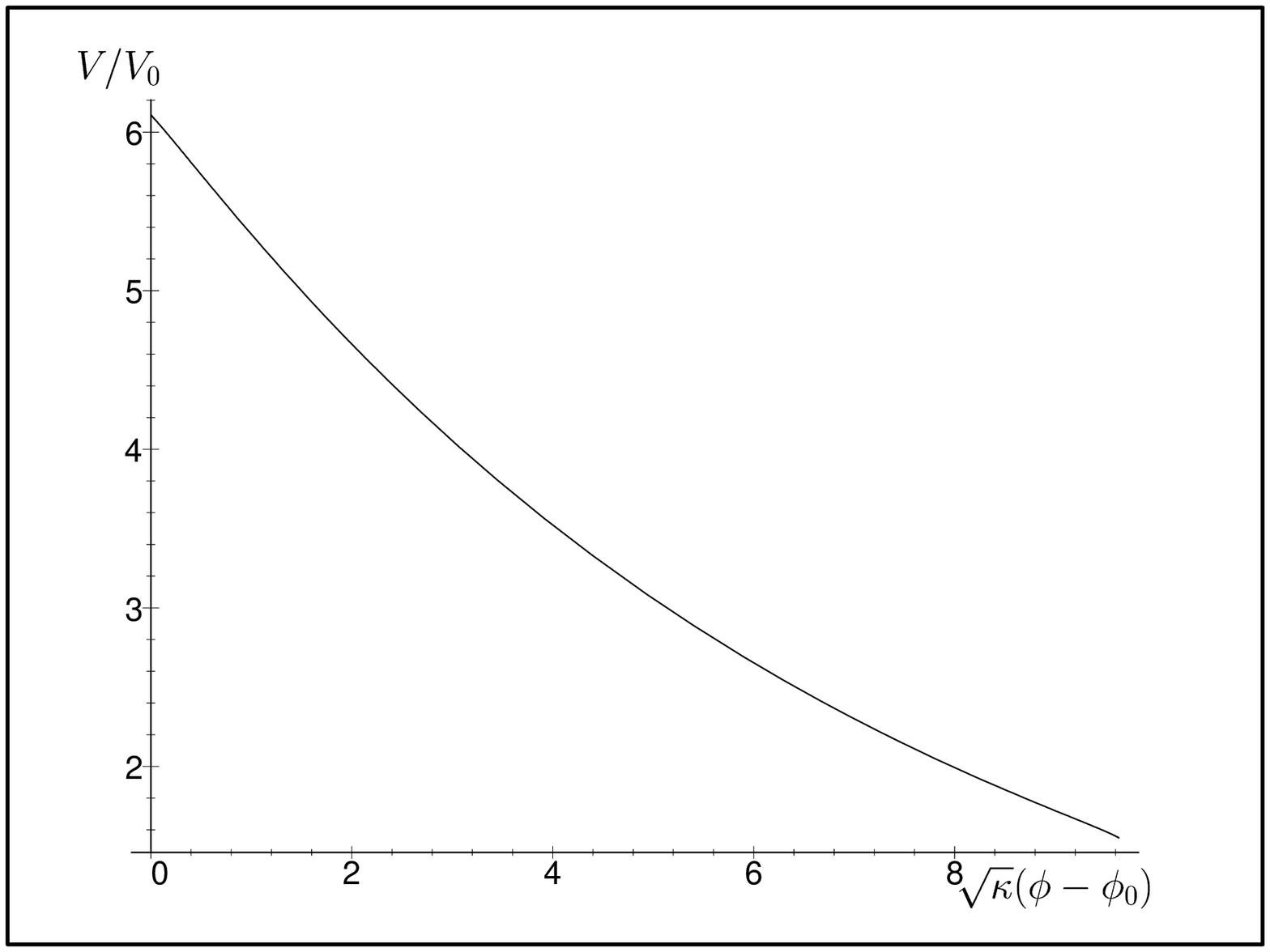,width=8.5cm}
\psfig{file=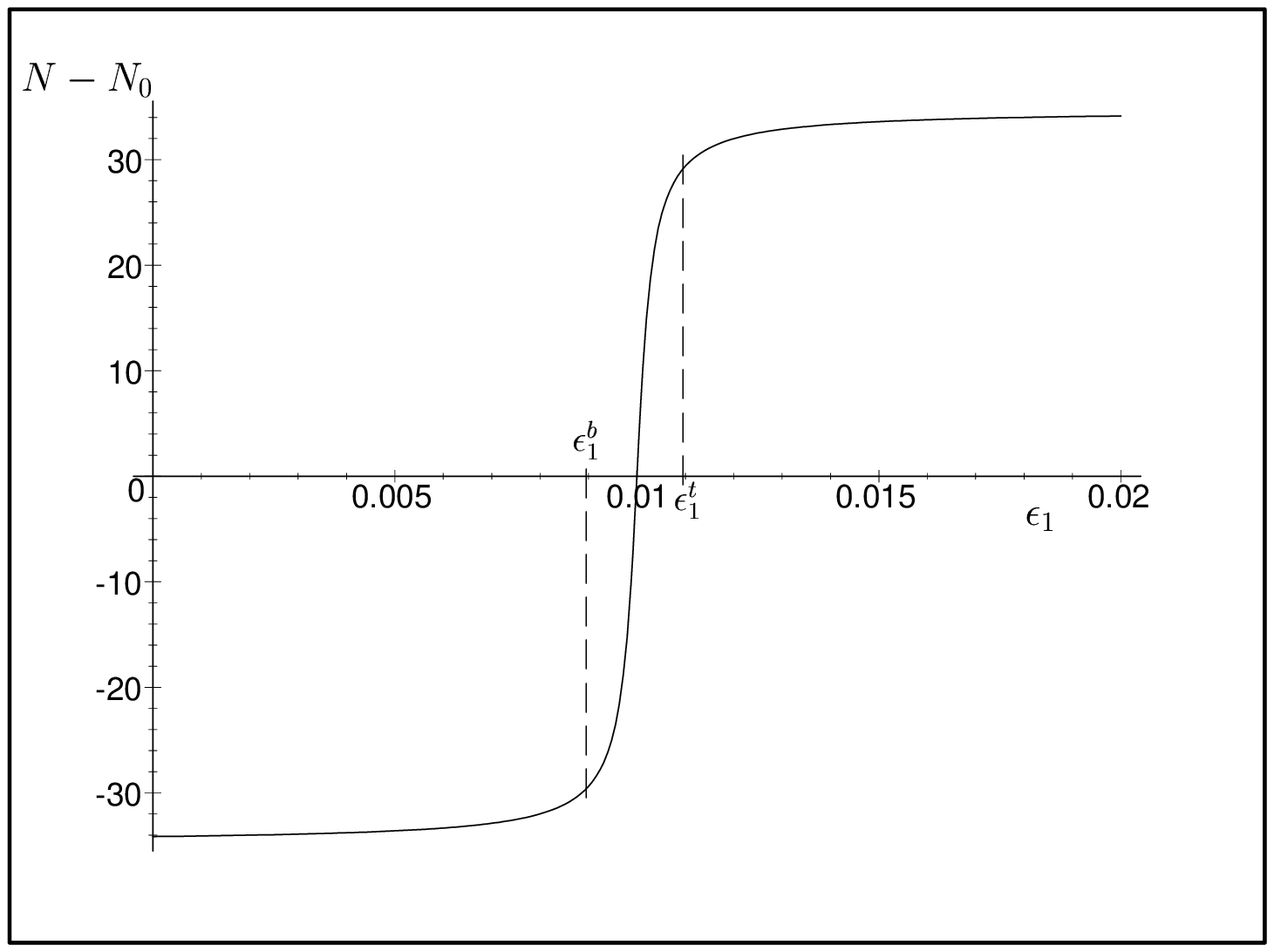,width=8.5cm}}
\caption{Case 2. Plots of the potential as function of $\phi$ (left) and the
efolds number as function of $\epsilon_1$ (right).}
\label{fig:C2VpDNe}
\end{figure}

The corresponding parametric potential is,
\begin{equation}
V(\phi)= \left\{
\begin{array}{rcl}
\phi(\epsilon_1)&=&\frac{(C+1)}{\sqrt{\kappa}}\frac{1}{\sqrt{a}\sqrt{4ac-b^2}}
\left\{\sqrt{2\sqrt{ac}-b}\left[
-\arctan\left(\frac{2\sqrt{a\epsilon_1}+\sqrt{2\sqrt{ac}-b}}
{\sqrt{2\sqrt{ac}+b}}\right) \right. \right. \nonumber \\
&+& \left. \left. 
\arctan\left(\frac{2\sqrt{a\epsilon_1}-\sqrt{2\sqrt{ac}-b}}
{\sqrt{2\sqrt{ac}+b}}\right) \right]
+ \frac{\sqrt{2\sqrt{ac}+b}}2
\ln\left|\frac{\sqrt{a}\epsilon_1+\sqrt{c}+\sqrt{2\sqrt{ac}-b}}
{\sqrt{a}\epsilon_1+\sqrt{c}-\sqrt{2\sqrt{ac}-b}}\right|\right\}
+ \phi_0 \, ,  \\
&& \\
V(\epsilon_1)&=&V_0(3-\epsilon_1)
\left|a\epsilon_1^{2}+b\epsilon_1 +c\right|^{-\frac{C+1}a}
\exp\left[\frac{2(C+1)b}{a\sqrt{4ac-b^2}}
\arctan\left(\frac{2a\epsilon_1+b}{\sqrt{4ac-b^2}}\right)\right]
\, .
\end{array}
\right.
\label{eq:C2Vp}
\end{equation} 
Analyzing the behaviors of $\phi(\epsilon_1)$ and $V(\epsilon_1)$ 
plotted in Fig.~\ref{fig:C2pVe},
and the dynamics of $\epsilon_1$ (Fig.~\ref{fig:C2et}), it is observed that 
potential (\ref{eq:C2Vp}) plotted in the left part of
Fig.~\ref{fig:C2VpDNe} 
fulfills criterion (\ref{eq:epsConds}) for any $\epsilon_1\in[0,1)$.
Moreover, the integration of Eq.~(\ref{eq:DNintde}) for this case yields,
\begin{equation}
\label{eq:C2DNe}
N=2\frac{(C+1)}{\sqrt{4ac-b^2}}
\arctan\left(\frac{2a\epsilon_1}{\sqrt{4ac-b^2}}\right) + N_0
\, ,
\end{equation}
and from the plot presented in the right part of 
Fig.~\ref{fig:C2VpDNe} one can see that, with an appropriate selection
of $a$, $b$ and $c$, this scenario will produce the required amount of 
inflation.

Even if in this model the inflationary period can be finished by $\epsilon_1$
reaching unity (in fact $\epsilon_1=1$ is reached shortly after the power-law 
like period), since this potential does not have a minimum, we still need to 
appeal to the hybrid inflation mechanism in order to account for the big bang
relativistic matter. The triggering value could be any value of $\phi$
corresponding to $1\geq\epsilon_1>\epsilon_1^t$. The scales crossing the 
horizon 
after $\epsilon_1=\epsilon_1^t$ will be extraordinarily small and reenter 
it back immediately with not relevant effect.

In this scenario, the values of possible initial conditions for the inflaton 
roll-down can be
divided in three intervals, namely, $L_1\equiv[0,\epsilon_1^b]$, 
$L_2$ and 
$L_3\equiv[\epsilon_1^t,1)$. Starting from $L_3$ negligible
expansion will be produced and the universes will collapse almost 
immediately. Since in $L_2$ $\epsilon_1$ is almost constant, $L_2$ is
a very narrow interval and the initial
conditions are only slightly different. Nevertheless, even that tiny difference
could account for very different amount of inflation and, correspondingly, to
universes with different degree of flatness, homogeneity and so on. Finally,
any inflationary dynamics starting from $L_1$ will face the huge expansion
produced in $L_2$ and the resulting perturbations spectra will be very similar
each other, differing from a power-law spectrum only at very large and very 
small scales.

Putting $H^2$ as given by Eq.~(\ref{eq:C2te}) in condition 
(\ref{eq:Cond4EtInf}) yields,
\begin{equation}
\frac{H_0^2}{\pi m_{Pl}^2 \epsilon_{1i}}
\left|a\epsilon_{1i}^2+b\epsilon_{1i}+c\right|^{-\frac{C+1}{a}}
\exp\left[\frac{2(C+1)b}{a\sqrt{4ac-b^2}}
\arctan\left(\frac{2a\epsilon_{1i}+b}{\sqrt{4ac-b^2}}\right)\right] 
\geq 1\, .
\end{equation}
For any $H_0$ there is a sufficiently small $\epsilon_{1i}$ such that
condition (\ref{eq:Cond4EtInf}) will be met, meaning that there is a 
larger number of
Hubble regions where the initial conditions for the scalar field roll-down
toward the big bang will belong to $L_1$. It implies that, for
inflation driven by potential (\ref{eq:C2Vp}), even if there are finite 
probabilities
for several different inflationary outputs, the most likely one is that of 
universes 
expanded by the same factor given by $L_2$ and with power-law like spectra of 
perturbations in a wide range of scales.

\section{Conclusions}
\label{sec:Concl}
Two inflationary scenarios were introduced with weakly scale-dependent spectral
indices yielding perturbations amplitudes which, within current observational
limits, can be fairly approximated by power-law spectra in a wide range 
of angular scales.  

If, as it could be expected, any of these potentials strongly resembles the 
actual inflaton potential, then the big bang universes arising in the
eternal inflation picture will be very similar each to the other from the
observational point of view. In that case, our observable Universe will be
a generic outcome of the cosmological evolution rather than an extraordinary
entity in the multiverses space.

This conclusion is limited by the assumptions underlying the calculations. 
First of all, the precision of the expressions for the spectral indices upon
which the SLIP is based might determine the functional forms of the spectral
indices and, consequently, of the corresponding inflaton potential. Since the 
perturbations spectra is expected to be nearly power-law, the information added
by higher order corrections may be irrelevant though that must be proved. Work
in this direction is in progress.

On the other hand, our predictions are valid for inflation taking place in the 
neighborhood of a given vacuum state. The observable size of this neighborhood 
is 
determined by the range of scales probed by the CMB experiments and, obviously,
will always be finite. Nevertheless, a theory may have several equivalent vacua
arising through different fundamental symmetries breaking. In this case, 
inflation can happen around every vacuum and, if the vacua are not symmetric, 
it leads to different big bang 
processes, this way diversifying the cosmological outcomes. The observation of
our Universe will not yield direct information on the 
inflationary dynamics taking place near other vacua. However, it could be 
expected that our conclusion holds for all of the universes related with each 
vacuum. Therefore, the universes would be grouped in a number of classes 
corresponding to the number of actually different vacuum states. Whether 
universes like
our are a generic cosmological product will depend on how large could this 
number be.

Finally, our conclusion heavily relies on the fact that most initial conditions
are set in the higher energies regime. If the characteristic energy scale is
too close or beyond the Planck scale, the validity of the semi-classical
analysis can break down.  
  
\acknowledgements
We are grateful to J.~E.~Lidsey, E.~J.~Copeland and R.~Abramo for useful 
discussions. 
This work is supported in part by the CONACyT 
grant 38495--E and the Sistema Nac. de Investigadores (SNI).

\end{document}